\documentclass[prb,twocolumn,showpacs,floatfix,preprintnumbers,amsmath,amssymb,superscriptaddress]{revtex4}

\usepackage{graphicx}
\usepackage{amsmath}
\usepackage{amssymb}
\usepackage{color}
\usepackage{dcolumn}
\usepackage{epsfig}
\usepackage{bm}
\usepackage{subfigure}
\usepackage[urlcolor=blue]{hyperref}
\hypersetup{backref, colorlinks=true, linkcolor=blue, citecolor=blue}
\begin{document}

\title{Pressure-induced superconductivity in palladium sulfide}

\author{Liu-Cheng Chen}
\affiliation{Institute of Solid State Physics, Chinese Academy of Sciences, Hefei 230000, China}
\affiliation{University of Science and Technology of China, Hefei 230026, China}
\affiliation{Center for High Pressure Science and Technology Advanced Research, Shanghai 201203, China}

\author{Hao Yu}
\affiliation{Center for High Pressure Science and Technology Advanced Research, Shanghai 201203, China}

\author{Hong-Jie Pang}
\affiliation{Center for High Pressure Science and Technology Advanced Research, Shanghai 201203, China}

\author{Bin-Bin Jiang}
\affiliation{State Key Laboratory of High Performance Ceramics and Superfine Microstructure, Shanghai Institute of Ceramics, Chinese Academy of Science, Shanghai 200050, China}
\affiliation{University of Chinese Academy of Sciences, Beijing 100049, China}

\author{Lei Su}
\affiliation{Key laboratory of Photochemistry, Institute of Chemistry, University of Chinese Academy of Sciences, Chinese Academy of Sciences, Beijing 100190, China}

\author{Xun Shi}
\affiliation{State Key Laboratory of High Performance Ceramics and Superfine Microstructure, Shanghai Institute of Ceramics, Chinese Academy of Science, Shanghai 200050, China}

\author{Li-Dong Chen}
\affiliation{State Key Laboratory of High Performance Ceramics and Superfine Microstructure, Shanghai Institute of Ceramics, Chinese Academy of Science, Shanghai 200050, China}

\author{Xiao-Jia Chen}
\email{xjchen@hpstar.ac.cn}
\affiliation{Center for High Pressure Science and Technology Advanced Research, Shanghai 201203, China}

\date{\today}

\begin{abstract}

An extended study on PdS is carried out with the measurements of the resistivity, Hall coefficient, Raman scattering, and X-ray diffraction at high pressures up to 42.3 GPa. With increasing pressure, superconductivity is observed accompanying with a structural phase transition at around 19.5 GPa. The coexistence of semiconducting and metallic phases observed at normal state is examined by the Raman scattering and X-ray diffraction between 19.5 and 29.5 GPa. After that, only the metallic normal state maintains with an almost constant superconducting transition temperature. The similar evolution between the superconducting transition temperature and carrier concentration with pressure supports the phonon-mediated superconductivity in this material. These results highlight the important role of pressure played in inducing superconductivity from these narrow band-gap semiconductors.

\end{abstract}


\maketitle

\section{INTRODUCTION}

Since the discovery of LaFeAsO$_{1-x}$F$_x$ superconductor, various Fe-based superconductors have been synthesized with the motivation to enhance superconducting transition temperature ($T_c$).\cite{kamih,xhchen,hlshi,jeevan} Among them, iron-based chalcogenides have been of great interest because of their fascinating superconducting properties.\cite{glasb,ferna,mahmou,qswang,qswang1} For example, FeSe is regarded as a reference material to elucidate the physics of high $T_c$ superconductivity in Fe-based compounds, due to the simplest crystallographic structure.\cite{cwluo,pdiko,fwang,hycao} Pressure is a clean tool to find new superconductors or enhance the properties of existing superconductors without inducing impurities. The $T_c$ of FeSe is 8 K at ambient pressure, and increases to 36.7 K under external pressure.\cite{fchsu,medve} Although FeSe system possesses many attractive features, lots of physical properties still remain unsettled. For example, FeSe$_{1-x}$ always composes of primarily PbO-type tetragonal phase with space group pf $P4$/$nmm$ and NiAs-type hexagonal phase with space group of $P6_{3}$/$mmc$. \cite{fchsu} At low temperatures, a structural transition from $P4$/$nmm$ to an orthorhombic phase with space group was reported from neutron diffraction measurements.\cite{mill} The central issue regarding which phase(s) accounts for superconductivity remains unknown. At high pressures, the $T_{c}$ evolution paths detected from different groups are very different.\cite{medve,garb,tiss,brai,miyo} There has been no agreement about the high-pressure structures and phase transformation for this so-called simple material.\cite{medve,garb,marg,kumar} It has been generally accepted\cite{imai} that applied pressure enhances spin fluctuations and thus $T_{c}$ in Fe$_{1+y}$Se, suggesting a close linkage between spin fluctuations and the mechanism of superconductivity. This seemingly beautiful picture does not get further support from the direct dection of magnetic excitations from neutron scattering measurements.\cite{mart} The spin resonance energy was not observed to be proportional to $T_{c}$ at high pressures.\cite{mart} This indicated that paring beyond magnetic excitations should account or at least work together with for superconductivity in FeSe. On the other hand, the strong pressure-dependent electron-phonon coupling was suggested from theoretical study on FeSe.\cite{mand} The newly developed techniques based on photoemission and x-ray free electron laser were used to determine the weight of interactions. The new experiment confirmed the theoretically predicted enhancement of the electron-phonon coupling strength owning to electron-electron correction effects.\cite{gerb} The sister compound FeS has also attracted great interest due to the comparability with FeSe.\cite{ajdevey,kdkwon} Superconductivity in tetragonal FeS was observed with a $T_c$ of 5 K, but $T_c$ gradually decreases with pressure, contrary to the case in FeSe.\cite{xflai,xflai1} Although there has been no report of superconductivity in the undoped FeTe, the extensive studies on FeTe system have promoted the understanding of superconducting mechanism in Fe-based superconductors.\cite{mizug,subed,mahes} In this approach, it is quite natural to pay attention to other analogous chalcogenides without iron.

\begin{figure*}[tbp]
\includegraphics[width=1\textwidth]{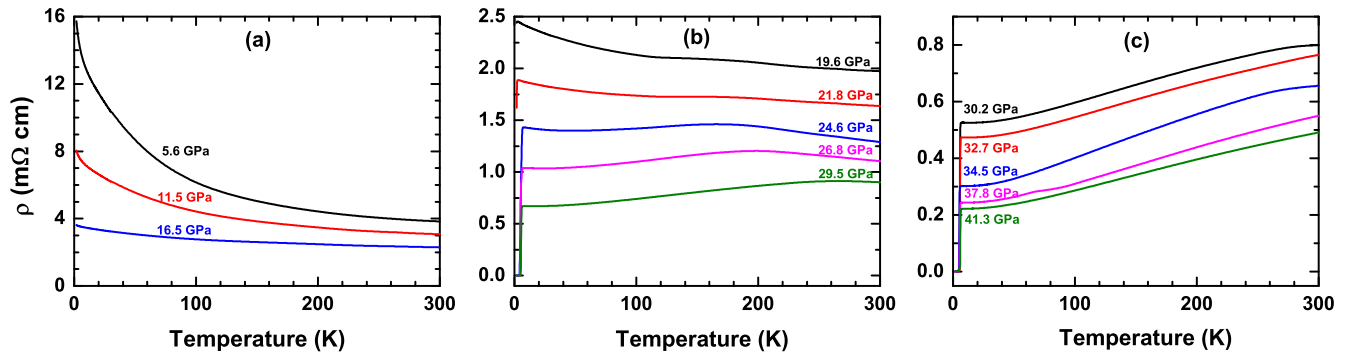}
\caption{Temperature dependence of the resistivity of PdS at various pressures. (a) The resistivity is gradually suppressed with increasing pressure and behaves like a typical semiconductor. (b) The metallic behaviour gradually becomes the dominant with increasing pressure at normal state and superconductivity emerges at 19.6 GPa. (c) The metallic state is fully occupied the crystal at normal state above 30 GPa and superconductivity can be seen obviously.}
\end{figure*}

To simplify the potential complex in understanding superconductivity in transition metal chalcogenides, one expects to explore superconductivity in other chalcogenides without magnetic elements. Palladium chalcogenides may offer alternative route for such a purpose.  For example,  PdTe was found to be a superconductor with a strongly coupled character.\cite{karki} By the application of pressure, $T_c$ was theoretically predicated to decrease in this compound.\cite{jchen} The high-pressure experiments have not  been carried out on PdTe yet. However, it is interesting to study the high-pressure behaviors and the evolution of superconductivity with pressure in palladium chalcogenides. So far, many physical properties of this family at high pressures have not been investigated experimentally. Palladium sulfide (PdS) has a wealth of superior physical properties, such as semiconducting, photoelectrochemical, and photovoltaic properties.\cite{ferre,jcwfo,baraw} These properties are being developed for device applications in catalysis and acid resistant high-temperature electrodes.\cite{blado,palla,chyan,zubko} Recently, this material was identified to be a potential thermoelectric material with large Seebeck coefficient and high electrical conductivity.\cite{liucheng} Pressure has often been used to drive narrow band semiconductors to become superconductors.\cite{ying,xjc} It is highly desired to examine whether superconductivity can be realized in PdS upon lattice compression and what the path will be undertaken at high pressures.

\begin{figure}[tbp]
\includegraphics[width=0.42\textwidth]{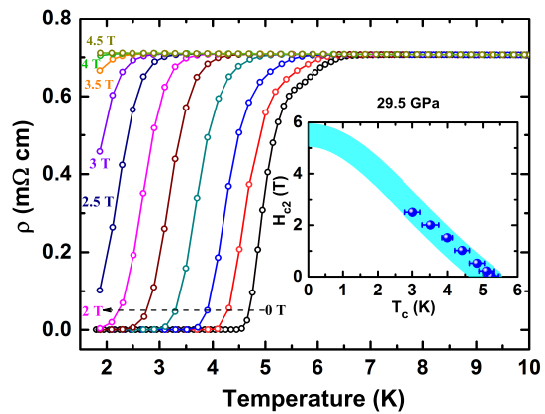}
\caption{Temperature dependence of the resistivity of PdS at a pressure of 29.8 GPa with applied magnetic fields. Inset: Upper critical field $H_{c2}$ at the pressure of 29.8 GPa. The color area represents the calculated $H_{c2}$ from the Werthamer-Helfand-Hohenberg equation.}
\end{figure}

To address the issues mentioned above, we perform a series of measurements to study the physical properties in PdS under pressure up to 42.3 GPa. We find that the semiconducting behavior is strongly suppressed and a metallic phase develops under pressure from the resistivity measurements. A pressure-induced superconductivity is observed when the phase transition occurs. The phase transition is confirmed by Raman scattering and X-ray diffraction measurements. The high-pressure phase diagram is thus established for this interesting material.

\begin{figure*}[tbp]
\includegraphics[width=0.9\textwidth]{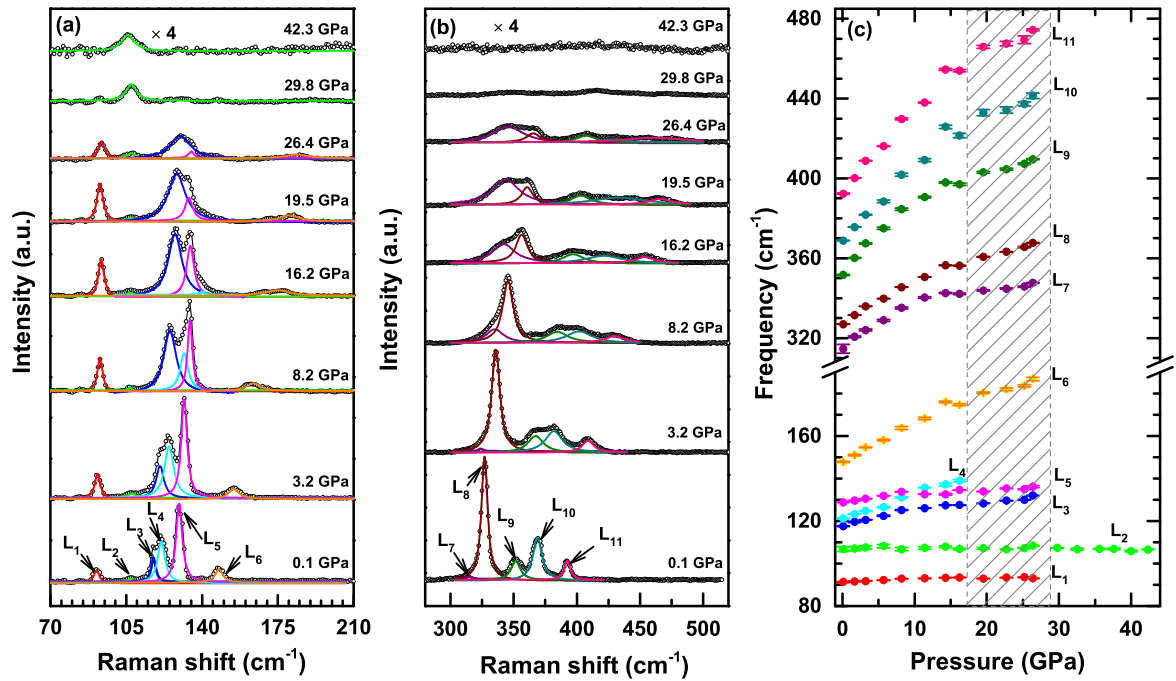}
\caption{Selected Raman spectra of PdS at room temperature and various pressures in the ranges of (a) 70 - 210 cm$^{-1}$ and (b) 280 - 520 cm$^{-1}$. (c) Pressure dependence of the obtained phonon frequencies. The dashed region is transition area and represents the coexistence of two phases.}
\end{figure*}

\section{EXPERIMENTAL DETAILS}

The high-quality sample PdS used in this experiment was synthesized by melting-quenched method and spark plasma sintering, detailed elsewhere.\cite{xyli} A nonmagnetic diamond anvil cell (DAC) made of Cu-Be alloy was used for high-pressure resistivity measurements.\cite{gavri} This customized cell has two symmetrical anvil culets with the diameter of 200 $\mu$m. An insulated rhenium flake was used as the gasket with the sample chamber of diameter 100 $\mu$m. The high-quality PdS was pressed into powders before filled in the sample chamber. A standard four-probe method with four Pt wires linked the sample powers and external Cu wires was used for the resistivity measurements in Physical Properties Measurement System (PPMS). For the Raman scattering experiments, pressure was realized by a symmetrical DAC with a 300 $\mu$m culet and the diameter of sample champer was 150 $\mu$m. The sample powers were loaded in the hole with the same conditions for the resistivity measurements. The power of exciting laser was 2 $mW$ with a wavelength of 488 nm, and the integral time was 5 mins in order to obtain better Raman scattering spectra. The high pressure X-ray scattering diffraction experiments, prepared the same environment as Raman measurements, were conducted at the Shanghai Synchrotron Radiation Facility with a wavelength 0.6199 {\AA}. In all the experiments mentioned above, pressure was calibrated by using the ruby fluorescence shift,\cite{hkmao} and the pressure was implemented around room temperature.

\section{RESULTS AND DISCUSSION}

Figure 1 shows the temperature dependence of the resistivity for PdS at various pressures up to 41.3 GPa. From Fig. 1(a), we can see that the resistivity of PdS is strongly suppressed with increasing pressure, and a typical semiconducting feature is observed from the resistivity behavior in all the temperature range at various pressures up to 16.5 GPa. Interestingly, when pressure is increased to 19.6 GPa, it enters into a superconducting state at low temperatures [Fig. 1(b)]. With further increasing pressure, the resistivity is gradually reduced and the superconductivity becomes more evident. Moreover, a resistivity hump is observed around 170 K at the pressure of 19.6 GPa. At the same time, a metallic behavior emerges at the normal state of this superconductor below the hump-emerging temperature. This phenomenon indicates an incomplete phase transition from a semiconductor state to a metal state induced by pressure above 19.6 GPa. The origin of the resistivity hump shifting towards to higher temperatures with increasing pressure is derived from the competing of these two phases. When the pressure is increased to 30.2 GPa, the semiconducting state is changed to the metallic state for the normal state. At 41.3 GPa, the highest pressure studied in this article, superconductivity is still persisting with an improved $T_c$ [Fig. 1(c)]. These results suggest that a new superconductor is discovered at high pressures, which may be accompanied by the structural phase transition.

\begin{figure}[tbp]
\includegraphics[width=0.44\textwidth]{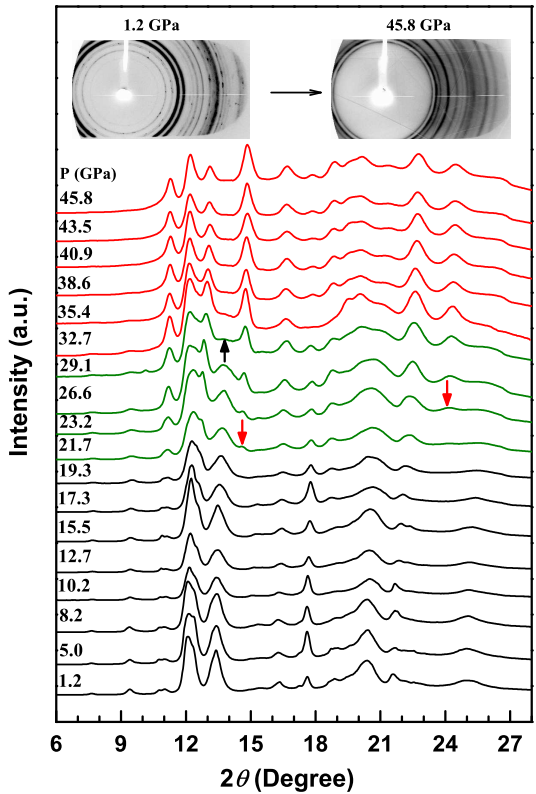}
\caption{High-pressure synchrotron X-ray diffraction patterns of PdS. The disappeared and emerging peaks are marked by black and red arrows, respectively. The original diffraction patterns (1.2 GPa and 45.8 GPa) are shown on the top.}
\end{figure}

The emergence of superconductivity is confirmed by the measurements of temperature dependence of the resistivity at different magnetic fields. The results at the pressure of 29.5 GPa are summarized in Fig. 2. It can be seen that the temperature dependent resistivity curve gradually shifts towards the low temperatures with increasing magnetic fields. Finally, the the temperature dependent resistivity almost becomes flat between 1.8 and 10 K with the magnetic field of 4.5 Tesla, which indicates the complete suppression of superconductivity in PdS. These magnetic-field dependent measurements provide convincing evidence for the detected superconducting transition at high pressures. Within the weak-coupling Bardeen-Cooper-Schrieffer (BCS) theory, an upper critical field at T = 0 K can be determined by the Werthamer-Helfand-Hohenberg equation:\cite{BCS} $H_{c2}(0) = 0.693[-(dH_{c2}/dT)]_{T_c}T_c$. The calculated $H_{c2}(0)$ is about 5.5 Tesla. The colorful area shown in the inset of Fig. 2 is the fitted temperature dependence of $H_{c2}$ with the expression:$H_{c2}(T) = H_{c2}(0)[1-(T/T_c)^2]/[1+(T/T_c)^2]$ based on the Ginzburg-Landau theory.\cite{NNi}

In order to shed insight into the origin of superconductivity in PdS at high pressures, the Raman scattering measurements were performed at the same condition as resistivity measurements. As we known, Raman spectroscopy can provide valuable information on the lattice vibration. Thus, it is a powerful tool to probe the structural evolution and phase transition. The selected room temperature Raman spectra of PdS at various pressures up to 42.3 GPa are shown in Fig. 3(a) and 3(b). To clarity, the obtained Raman vibration modes are marked by arrows and denoted as L$_n$, respectively. Upon increasing pressure, most of the obtained Raman modes are broad at low pressures, indicating a strong electron-phonon coupling and phonon-phonon interaction. Above 16.2 GPa, the Raman mode L$_4$ is disappearing which indicates the appearance of a new high-pressure phase. In other words, the sample PdS has evolved into a coexistence of two phases including the initial phase (tetragonal structure) and a new phase (the metallic phase). This is consistent with the emerging of superconductivity from the resistivity measurements [Fig. 1(b)]. With further increasing pressure, the gradually weakened intensity of the Raman modes is consistent with the gradually obvious metallic normal state [Fig. 1(b)]. Above 29.8 GPa, almost all the Raman modes disappear excepting the very week and broad L$_2$, indicating that the sample PdS completely evolved into the metallic phase as shown in Fig. 1(b).

Depending on the Lorentzian fitting,\cite{ranja} the pressure dependence of the obtained phonon frequencies are shown in Fig. 3(c). It is obvious that PdS undergoes a phase transition in the intermediary areas and the dashed area in a rough range between 16.2 and 29.8 GPa represents the coexistence of two phases. Almost all the Raman modes shift toward higher frequencies obviously under pressure before 19.6 GPa, originating from the expected contraction of interatomic distances. However, all the vibrational mode do not change too much between the coexisted phases. Above 26.5 GPa, only the vibrational mode L$_2$ ($\times$4) is still existing. Thus, the vibrational mode L$_2$ may be special to the strong electron-phonon coupling in favor of superconductivity.

Generally speaking, a sharp reduction of Raman peaks can be attributed to a structure transton with high symmetry where no Raman mode is allowed, or because of a transition to a metalic state which will shield the Raman signal due to the limited penetration depth of the exciting laser. In order to give a well-established evidence to the structure evolution in PdS, room-temperature synchrotron X-ray diffraction patterns of PdS at various pressures are measured up to 45.8 GPa. As shown in Fig. 4, all the Bragg peaks shift to larger angles with the increase of pressure, showing the shrinkage of the PdS lattice. Almost all diffraction peaks can be fitted to the tetragonal symmetry at low pressure, suggesting that the crystal structure of the PdS is stable below 19 GPa. Upon further compression, a new peak marked by a red arrow appears at 21.7 GPa, which means that a new phase emerges. Then, another new peak which is also marked by a red arrow appears at 23.2 GPa.  And these two new peaks become more evident with further increasing pressure. On the contrary, the Bragg peak marked by a black arrow disappears at the pressure of 29.1 GPa. These results suggest that the two phases coexist between 21.7 and 29.1 GPa, which is consistent with the resistivity and Raman scattering measurements. Above 32.7 GPa, the material completely evolutes into the new phase revealed by Raman spectra. The new phase is a more complex phase, which is also supported by the original diffraction patterns on the top of Fig. 4. The details of the new structure are still under investigation. However, it is certain that pressure drives PdS to evolve into a more complex structural phase above 19.5 GPa.

\begin{figure}[tbp]
\includegraphics[width=0.48\textwidth]{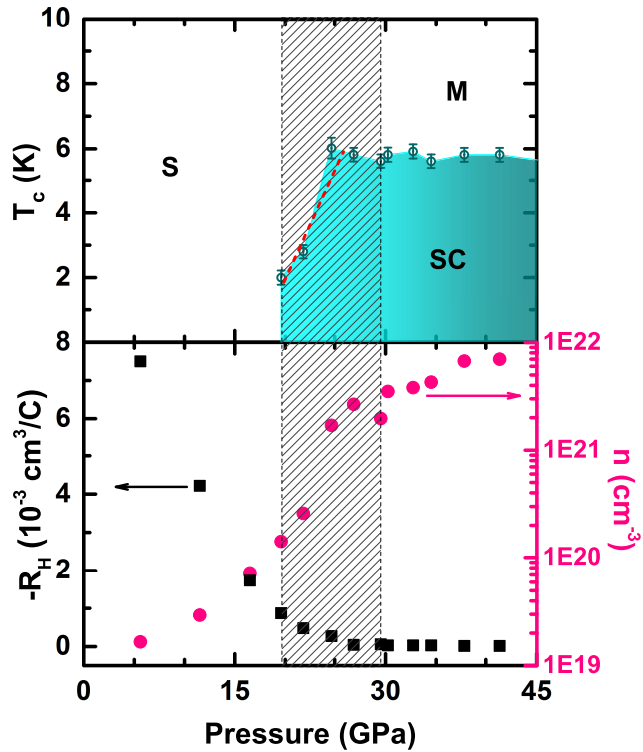}
\caption{Phase diagram of PdS at high pressures. The red-dot line is a linear fitting for d$T_c$/dP. The lower panel shows the pressure dependence of the Hall coefficient and carrier concentration measured at 10 K. The dashed region represent the phase boarders of a semiconductor (S) and metal (M). The dashed area represents the coexistence of two phases.}
\end{figure}

The phase diagram of PdS is mapped out by combining the resistivity, Raman scatting, and X-ray diffraction measurements (Fig. 5). It can be seen that PdS keeps it's initial phase with the behavior as a semiconductor (S) under the pressure of 19.5 GPa. Then, with increasing pressure, it evolves into a superconducting state with the coexistence of two phases between 19.5 and 29.8 GPa. It is specially interesting that $T_c$ has an obvious improvement in the region of the coexisted phases. With a roughly linear fitting (red-dot line), the initial pressure dependent coefficient of $T_c$, d$T_c$/dP ($\thicksim$ 0.58 K/GPa) is obtained. After the obvious improvement, $T_c$ is observed to level off with increasing pressure together with a complete evolution into a metallic (M) state at normal state. Overall, $T_c$ first exhibits a arc-like shape with increasing pressure and then becomes almost constant upon heavy compression. This behavior is very similar to a reported superconductor (NbN), where $T_c$ roughly increases with applied pressure up to 4 GPa, and then reaches a constant from 4 GPa to 42 GPa.\cite{xiaojc} As shown in the lower panel of Fig. 5, the negative Hall coefficient demonstrates that the dominant electron carrier character for the superconducting phase. The sign of the calculated carrier concentration at 10 K gradually increases with increasing pressure in initial phase region. Then, the carrier concentration has a relatively weak pressure dependence at higher pressures after a noticeable rise in the region of the coexisted phases. The carrier concentration for the superconducting phase is in the range between 10$^{20}$ and 10$^{22}$ cm$^{-3}$, at least two order higher in magnitude than that at ambient condition, which is very beneficial to the electron-phonon coupling in favor of superconductivity. The pressure dependent $T_c$ and evolution of carrier concentration under pressure are coherent, especially the obvious increase of them around the same pressure range. This phenomenon may have provided a natural explanation for the superconductivity of PdS under pressure.

The properties of conventional superconductors can be explained well by the theory of BCS, and the superconductivity is always driven from the coupling between electrons and phonons (lattice vibrations).\cite{bcsll} Recently, the strong electron-phonon interaction has been observed in transition metal chalcogenides, which indicates the importance of electron-phonon coupling for high temperature superconductivity.\cite{mand,gerb,Lanzara,Koufos} PdS as a transition metal chalcogenide, should be also an electron-phonon coupled superconductor.

For phonon-mediated superconductivity, $T_c$ can be expressed as $T_c$$\propto$$\langle\omega_{log}\rangle$exp$[-1.04(1+\lambda)/\lambda]$.\cite{wlmci,eaeki} Here, the electron-phonon coupling constant $\lambda$ is determined by the equation of $\lambda$=$\frac{N(0)\langle I^2\rangle}{M\langle\omega^2\rangle}$, where $\langle\omega_{log}\rangle$ is phonon energy, $N(0)$ is the density of electronic states at the Fermi energy, $\langle I^2\rangle$ is the mean-square electron-phonon matrix element, $M$ is the ionic mass, and $\langle\omega^2\rangle$ is the mean-square frequencies over the phonon spectra. Previous studies have established that $\langle I^2\rangle$ and $M$ do not change too much with pressure.\cite{xiaojc,xjc2} The information about phonon energy could be estimated from the Raman spectra (Fig. 3). In the superconducting regime, the phonon energy $\langle\omega_{log}\rangle$ and $\langle\omega^2\rangle$ are not sensitive to pressure. The electronic part of $\lambda$ could be reflected by carrier concentration which is proximate to $N(0)$ (Fig. 5). It is obvious that the pressure dependence $T_c$ and carrier concentration is synchronous in the superconducting regime. Combined with the relationship of $T_c$ and $\lambda$, there is no doubt that the electron-phonon pairing mechanism plays an important role in superconductivity of PdS.

\section{Conclusions}

In summary, we have synthesized high-quality sample PdS with single phase at ambient pressure. High-pressure resistivity, Hall coefficient, Raman scattering, and X-ray diffraction measurements have been performed on PdS up to 42.3 GPa. With increasing pressure, the semiconducting behavior has been strongly suppressed. A metallic phase has emerged and gradually occupied the dominant position upon further compression. The pressure-driven semiconductor-metal transition has been examined by combining Raman scattering and X-ray diffraction. Superconductivity has been observed when the metallic phase emerges at pressure of 19.6 GPa. The character of superconductivity has been confirmed by the obtained zero resistance and the suppression of the resistivity with the application of magnetic fields. We suggested that the evolution of $T_c$ under pressure was dominated by the carrier concentration. These findings enrich the superconducting family from transition metal chalcogenides. PdS-derived superconductors with higher $T_c$ are expected under pressure. These results will shed new light on understanding the mechanism of superconductivity in this kind of materials.

\begin{acknowledgments}
Lei Su acknowledged the support from the Natural Science Foundation of China (No. 21273206). Xun Shi and Li-Dong Chen acknowledged the support from the National Basic Research Program of China (973-program) under Project No. 2013CB632501, the Natural Science Foundation of China under the No. 11234012, and the Shanghai Government (Grant No.15JC1400301).
\end{acknowledgments}


\begin{references}

\bibitem{kamih}Y. Kamihara, T. Watanabe, M. Hirano, and H. Hosono, Iron-based layered superconductor La[O$_{1-x}$F$_x$]FeAs (x=0.05-0.12) with $T_c$ = 26 K, J. Am. Chem. Soc. \textbf{130}, 3296 (2008).

\bibitem{xhchen}X. H. Chen, T. Wu, G. Wu, R. H. Liu, H. Chen, and D. F. Fang, Superconductivity at 43 K in SmFeAsO$_{1-x}$F$_x$, Nature \textbf{453}, 761 (2008).

\bibitem{hlshi}H. L. Shi, H. X. Yang, H. F. Tian, J. B. Lu, Z. W. Wang, Y. B. Qin, Y. J. Song, and J. Q. Li, Structural properties and superconductivity of SrFe$_2$As$_{2-x}$P$_x$(0.0 $\leq$ x $\leq$ 1.0) and CaFe$_2$As$_{2-y}$P$_y$ (0.0 $\leq$ y $\leq$ 0.3), J. Phys: Condens. Mat. \textbf{22}, 1257002 (2010).

\bibitem{jeevan}H. S. Jeevan, Z. Hossain, Deepa Kasinathan, H. Rosner, C. Geibel, and P. Gegenwart, High-temperature superconductivity in Eu$_{0.5}$K$_{0.5}$Fe$_2$As$_2$, Phys. Rev. B \textbf{78}, 092406 (2008).

\bibitem{glasb}J. K. Glasbrenner, I. I. Mazin, Harald O. Jeschke, P. J. Hirschfeld,	R. M. Fernandes, and R. Valenti, Effect of magnetic frustration on nematicity and superconductivity in iron chalcogenides, Nat. Phys. \textbf{11}, 953 (2015).

\bibitem{ferna}R. M. Fernandes, A. V. Chubukov, and J. Schmalian, What drives nematic order in iron-based superconductors? Nat. Phys. \textbf{10}, 97 (2014).

\bibitem{mahmou}M. Abdel-Hafiez, Y. Y. Zhang, Z. Y. Cao, C. G. Duan, G. Karapetrov, V. M. Pudalov, V. A. Vlasenko, A. V. Sadakov, D. A. Knyazev, T. A. Romanova, D. A. Chareev, O. S. Volkova, A. N. Vasiliev, and X. J. Chen, Superconducting properties of sulfur-doped iron selenide, Phys. Rev. B \textbf{91}, 165109 (2015).

\bibitem{qswang}Q. S. Wang, Y. Shen, B. Y. Pan, X. W. Zhang, K. Ikeuchi, K. Iida, A. D. Christianson, H. C. Walker, D. T. Adroja, M. Abdel-Hafiez, X. J. Chen, D. A. Chareev, A. N. Vasiliev, and J. Zhao, Magnetic ground state of FeSe, Nat. Commun. \textbf{7}, 12182 (2016).

\bibitem{qswang1}Q. S. Wang, Y. Shen, B. Y. Pan, Y. Q. Hao,	M. W. Ma, F. Zhou, P. Steffens,	K. Schmalzl, T. R. Forrest,	M. Abdel-Hafiez, X. J. Chen, D. A. Chareev,	A. N. Vasiliev,	P. Bourges,	Y. Sidis, H. B. Cao, and J. Zhao, Strong interplay between stripe spin fluctuations, nematicity and superconductivity in FeSe, Nat. Mater. \textbf{15}, 159 (2016).

\bibitem{cwluo}C. W. Luo, I. H. Wu, P. C. Cheng, J. Y. Lin, K. H. Wu, T. M. Uen, J. Y. Juang, T. Kobayashi, D. A. Chareev, O. S. Volkova, and A. N. Vasiliev, Quasiparticle dynamics and phonon softening in FeSe superconductors, Phys. Rev. Lett. \textbf{108}, 257006 (2012).

\bibitem{pdiko}P. Diko, V. Antal, V. Kavecansky, C. Yang, and I. Chen, Microstructure and phase transformations in FeSe superconductor, Physica C \textbf{476}, 29 (2012).

\bibitem{fwang}F. Wang, S. A. Kivelson, and D. H. Lee, Nematicity and quantum paramagnetism in FeSe. Nat. Phys. \textbf{11}, 959 (2015).

\bibitem{hycao}H. Y. Cao, S. Chen, H. Xiang, X. G. Gong, Antiferromagnetic ground state with pair-checkerboard order in FeSe. Phys. Rev. B \textbf{91}, 020504 (2015).

\bibitem{fchsu}F. C. Hsu, J. Y. Luo, K. W. Yeh, T. K. Chen, T. W. Huang, P. M. Wu, Y. C. Lee, Y. L. Huang, Y. Y. Chu, D. C. Yan, and M. K. Wu, Superconductivity in the PbO-type structure alpha-FeSe, Proc. Natl Acad. Sci. \textbf{105}, 14262 (2008).

\bibitem{medve}S. Medvedev, T. M. Mcqueen, I. A. Troyan, T. Palasyuk, M. I. Eremets, R. J. Cava, S. Naghavi, F. Casper, V. Ksenofontov, G. Wortmann, and C. Felser, Electronic and magnetic phase diagram of bold italic $\beta$-Fe$_{1.01}$Se with superconductivity at 36.7 K under pressure, Nat. Mater. \textbf{8}, 630 (2009).

\bibitem{ajdevey}A. J. Devey, R. Grau-Crespo, and N. H. de Leeuw, Combined density functional theory and interatomic potential study of the bulk and surface structures and properties of the iron sulfide mackinawite (FeS), J. Phys. Chem. C \textbf{29} 10960 (2008).

\bibitem{kdkwon}K. D. Kwon, K. Refson, S. Bone, R. M. Qiao, W. L. Yang, Z. Liu, and G. Sposito, Magnetic ordering in tetragonal FeS: Evidence for strong itinerant spin fluctuations, Phys. Rev. B \textbf{83}, 064402 (2011).

\bibitem{xflai}X. F. Lai, H. Zhang, Y. Q. Wang, X. Wang, X. Zhang, J. H. Lin, and F. Q. Huang, Observation of superconductivity in tetragonal FeS, J. Am. Chem. Soc. \textbf{32}, 10148 (2015).

\bibitem{xflai1}X. F. Lai, Y. Liu, X. J. L\"{u}, S. J. Zhang, K. J. Bu, C. Q. Jin, H. Zhang, J. H. Lin, and F. Q. Huang, Suppression of superconductivity and structural phase transitions under pressure in tetragonal FeS, Sci. Rep. \textbf{6}, 31077 (2016).

\bibitem{mizug}Y. Mizuguchi, F. Tomioka, S. Tsuda, T. Yamaguchi, and Y. Takano, FeTe as a candidate material for new iron-based superconductor, Physica C \textbf{469}, 1027 (2008).

\bibitem{subed}A. Subedi, L. J. Zhang, D. J. Singh, and M. H. Du, Density functional study of FeS, FeSe, and FeTe: Electronic structure, magnetism, phonons, and superconductivity, Phys. Rev. B \textbf{78}, 134514 (2008).

\bibitem{mahes}P. K. Maheshwari, R. Jha, B. Gahtori, and V. P. S. Awana, Structural and magnetic properties of flux-free large FeTe single crystal, J. Supercon. Nov. Magn. \textbf{28}, 2893 (2015).

\bibitem{mill} J. N. Millican, D. Phelan, E. L.Thomas, J. B. Le$\dot{a}$o, and E. Carpenter, Pressure-induced effects on the structure of the FeSe superconductor, Solid State Commun. \textbf{149}, 707 (2009).

\bibitem{garb} G. Garbarino, A. Sow, P. Lejay, A. Sulpice, P. Toulemonde, M. Mezouar, and M. N$\acute{u}\tilde{n}$ez-Regueiro, High-temperature superconductivity ($T_c$ onset at 34 K) in the high-pressure orthorhombic phase of FeSe, Europhys. Lett. \textbf{86}, 27001 (2009).

\bibitem{tiss} V. G. Tissen, E. G. Ponyatovsky, M. V. Nefedova, A. N. Titov, and V. V. Fedorenko, Effects of pressure-induced phase transitions on superconductivity in single-crystal Fe$_{1.02}$Se, Phys. Rev. B \textbf{80}, 092507 (2009).

\bibitem{brai} D. Braithwaite, B. Salce, G. Lapertot, F. Bourdarot, C. Marin, D. Aoki, and M. Hanfland, Superconducting and normal phases of FeSe single crystals at high pressure, J. Phys.: Condens. Matter \textbf{21}, 232202 (2009).

\bibitem{miyo} K. Miyoshi,  Y. Takaichi, E. Mutou, K. Fujiwara, and J. Takeuchi, Anomalous Pressure Dependence of the Superconducting Transition Temperature in FeSe$_{1-x}$ Studied by DC Magnetic Measurements, Jpn. J. Phys. Soc. \textbf{78}, 093703 (2009).

\bibitem{marg} S. Margadonna, Y. Takabayashi, Y. Ohishi, Y. Mizuguchi, Y. Takano, T. Kagayama, T. Nakagawa, M. Takata, and K. Prassides, Pressure evolution of the low-temperature crystal structure and bonding of the superconductor FeSe ($T_c$=37 K), Phys. Rev. B \textbf{80}, 064506 (2009).

\bibitem{kumar} R. S. Kumar, Y. Zhang, S. Sinogeikin, Y. Xiao, S. Kumar, P. Chow, A. L. Cornelius, and C. F. Chen, Crystal and electronic structure of FeSe at high pressure and low temperature, J. Phys. Chem. B \textbf{114}, 12597 (2010).

\bibitem{imai} T. Imai, K. Ahilan, F. L. Ning, T. M. McQueen, and R. J. Cava, Why does undoped FeSe become a high-$T_c$ superconductor under Pressure? Phys. Rev. Lett. \textbf{102}, 177005 (2009).

\bibitem{mart} K. Marty, A. D. Christianson, A. M. dos Santos, B. Sipos, K. Matsubayashi, Y. Uwatoko, J. A. Fernandez-Baca, C. A. Tulk, T. A. Maier, B. C. Sales, and M. D. Lumsden, Effect of pressure on the neutron spin resonance in the unconventional superconductor FeTe$_{0.6}$Se$_{0.4}$, Phys. Rev. B \textbf{86}, 220509 (2012).

\bibitem{mand} S. Mandal, R. E. Cohen, and K. Haule, Strong pressure-dependent electron-phonon coupling in FeSe, Phys. Rev. B {\bf 89}, 220502(R) (2014).

\bibitem{gerb} S. Gerber et al., Femtosecond electron-phonon lock-in by photoemission and x-ray free-electron laser, Science {\bf 357}, 71 (2017).

\bibitem{karki} A. B. Karki, D. A. Browne, S. Stadler, J. Li, and R. Jin, PdTe: a strongly coupled superconductor, J. Phys: Condens. Mat. \textbf{24}, 055701 (2012).

\bibitem{jchen}J. Chen and X. Wang, Superconductivity origin of PdTe and pressure effect: Insights from first-principles investigation, Solid. State. Sci. \textbf{52}, 23 (2016).

\bibitem{ferre}I. J. Ferrer, P. D. Chao, A. Pascual, and C. S\'{a}nchez, An investigation on palladium sulphide (PdS) thin films as a photovoltaic material, Thin Solid Films \textbf{515}, 5783 (2007).

\bibitem{jcwfo}J. C. W. Folmer, J. A. Turner, and B. A. Parkinson, Photoelectrochemical characterization of several semiconducting compounds of palladium with sulfur and/or phosphorus, J. Solid. State. Chem. \textbf{68}, 28 (1987).

\bibitem{baraw}M. Barawi, I. J. Ferrer, J. R. Ares, and C. S\'{a}cnchez, Hydrogen evolution using palladium sulfide (PdS) nanocorals as photoanodes in aqueous solution, Acs. Appl. Mater. Inter. \textbf{6}, 20544 (2014).

\bibitem{blado}J. J. Bladon, A. Lamola, F. W. Lytle, W. Sonnenberg, J. N. Robinson, and G. Philipose, A Palladium sulfide catalyst for electrolytic plating, J. Electrochem. Soc. \textbf{143}, 1206 (1996).

\bibitem{palla}J. M. Alan, G. R. Antonio, R. M. Inmaculada, and A. A. James, Palladium sulphide-a highly selective catalyst for the gas phase hydrogenation of alkynes to alkenes, J. Catal. \textbf{340}, 10 (2016).

\bibitem{chyan}C. H. Yang, Y. Y. Wang, C. C. Wan, and C. J. Chen, A search for the mechanism of direct copper plating via bridging ligands, J. Electrochem. Soc. \textbf{143}, 3521 (1996).

\bibitem{zubko}A. Zubkov, T. Fujino, N. Sato, and K. Yamada, Enthalpies of formation of the palladium sulphides, J. Chem. Thermodyn. \textbf{30}, 571 (1998).

\bibitem{liucheng} L. C. Chen, B. B. Jiang, H. Yu, H. J. Pang, L. Su, X. Shi, L. D. Chen, and X. J. Chen, Binary palladium sulfide: A potential base thermoelectric material (unpublished).

\bibitem{ying}J. J. Ying, V. V. Struzhkin, Z. Y. Cao, A. F. Goncharov, H. K. Mao, F. Chen, X. H. Chen, A. G. Gavriliuk, and X. J. Chen, Realization of insulating state and superconductivity in the rashba semiconductor BiTeCl, Phys. Rev. B \textbf{93}, 100504 (2016).

\bibitem{xjc} X. J. Chen, C. Zhang, Y. Meng, R. Q. Zhang, H. Q. Lin, V. V. Struzhkin, and H. K. Mao, $\beta$$-$Imma$-$sh phase transitions of germanium,  Phys. Rev. Lett. {\bf 106}, 135502 (2011).

\bibitem{xyli}X. Y. Li, D. Li, H. X. Xin, J. Zhang, C. J. Song, and X. Y. Qin, Effects of bismuth doping on the thermoelectric properties of Cu$_3$SbSe$_4$ at moderate temperatures, J. Alloy. Compd. \textbf{561}, 105 (2013).

\bibitem{gavri}A. G. Gavriliuk, A. A. Mironovich, and V. V. Struzhkin, Miniature diamond anvil cell for broad range of high pressure measurements, Rev. Sci. Instrum. \textbf{80}, 043906 (2009).

\bibitem{hkmao} H. K. Mao, P. M. Bell, J. W. Shaner, and D. J. Stembey, Specific volume measurements of Cu, Mo, Pd, and Ag and calibration of the ruby R$_1$ fluorescence pressure gauge from 0.06 to 1 Mbar, J. Appl. Phys. \textbf{49}, 3276 (1978).

\bibitem{BCS}N. R.Werthamer, E. Helfand, and P. C. Hohenberg, Temperature and purity dependence of the superconducting critical field, H$_{c2}$. III. electron spin and spin-orbit effects, Phys. Rev. \textbf{147}, 295 (1966).

\bibitem{NNi}N. Ni, M. E. Tillman, J. Q. Yan, A. Kracher, S. T. Hannahs, S. L. Bud'ko, and P. C. Canfield, Effects of Co substitution on thermodynamic and transport properties and anisotropic H$_{c2}$ in Ba(Fe$_{1-x}$Co$_x$)$_2$As$_2$ single crystals, Phys. Rev. B \textbf{78}, 214515 (2008).

\bibitem{ranja}R. K. Singh, S. N. Singh, B. P. Asthana, and C. M. Pathak, Deconvolution of lorentzian Raman linewidth: Techniques of polynomial fitting and extrapolation, J. Raman Spectrosc. \textbf{25}, 423 (1994).

\bibitem{xiaojc}X. J. Chen, V. V. Struzhkin, Z. G. Wu, R. E. Cohen, S. Kung, H. K. Mao, R. J. Hemley, and A. N. Christensen, Electronic stiffness of a superconducting niobium nitride single crystal under pressure, Phys. Rev. B \textbf{72}, 094514 (2005).

\bibitem{bcsll}J. Bardeen, L. N. Cooper, and J. R. Schrieffer, Theory of superconductivity, Phys. Rev. \textbf{108}, 1175 (1957).

\bibitem{Lanzara}A. Lanzara, P. V. Bogdanov, X. J. Zhou, S. A. Kellar, D. L. Feng, E. D. Lu, T. Yoshida, H. Eisaki, A. Fujimori, K. Kishio, J. I. Shimoyama, T. Noda, S. Uchida, Z. Hussain and Z. X. Shen
Evidence for ubiquitous strong electron¨Cphonon coupling in high-temperature superconductors, Nature \textbf{412}, 510 (2001).

\bibitem{Koufos}A. P. Koufos, D. A. Papaconstantopoulos, and M. J. Mehl, First-principles study of the electronic structure of iron-selenium: Implications for electron-phonon superconductivity, Phys. Rev. B \textbf{89}, 035150 (2014).

\bibitem{wlmci}W. L. McMillan, Transition temperature of strongly coupled superconductors, Phys. Rev. \textbf{167}, 331 (1968).

\bibitem{eaeki}E. A. Ekimov, V. A. Sidorov, E. D. Bauer, N. N. Mel'nik, N. J. Curro, J. D. Thompson, and S. M. Stishov, Superconductivity in diamond, Nat. Lett. \textbf{428}, 542 (2004).

\bibitem{xjc2}X. J. Chen, V. V. Struzhkin, S. Kung, H. K. Mao, R. J. Hemley, and A. N. Christensen, Pressure-induced phonon frequency shifts in transition-metal nitrides, Phys. Rev. B \textbf{70}, 014501 (2004).

\end{references}
\end{document}